\documentclass[10pt, twocolumn]{article}

\usepackage[utf8]{inputenc}
\usepackage[T1]{fontenc}
\usepackage{lmodern}
\usepackage{booktabs}
\usepackage{tabularx} 
\usepackage{listings}
\usepackage{algorithm}
\usepackage{algpseudocode}
\usepackage{hyperref}

\algrenewcommand\algorithmiccomment[1]{\hfill\textcolor{gray}{// #1}}
\algrenewcommand\algorithmicensure{\textbf{Ensure:}}
\algrenewcommand\algorithmicrequire{\textbf{Require:}}

\usepackage{geometry}
\usepackage{graphicx}
\usepackage{caption}
\usepackage{subcaption}

\usepackage{amsmath, amssymb, amsthm}

\usepackage[numbers]{natbib}
\usepackage{xcolor}
\usepackage{hyperref}
\hypersetup{
    colorlinks=true,
    linkcolor=blue,
    citecolor=blue,
    urlcolor=blue,
}

\usepackage{balance}
\title{Latency Based Tiling}
\author{
    Jack Cashman \\
    \texttt{jcashman@u.rochester.edu}
}
\date{}

\begin{document}

\twocolumn[
\maketitle
\begin{abstract}
\noindent
Latency Based Tiling provides a systems based approach to deriving approximate tiling solution that maximizes locality while maintaining a fast compile time. The method uses triangular loops to characterize miss ratio scaling of a machine avoiding prefetcher distortion \citet{sun2024}. Miss ratio scaling captures the relationship between data access latency and working set size with sharp increases in latency indicating the data footprint exceeds capacity from a cache level. Through these noticeable increases in latency we can determine an approximate location for L1, L2, and L3 memory sizes. These sizes are expected to be under approximations of a systems true memory sizes which is in line with our expectations given the shared nature of cache in a multi process system as described in defensive loop tiling \citet{bao2013}. Unlike auto tuning, which can be effective but prohibitively slow, Latency Based Tiling achieves negligible compile time overhead. The implementation in Rust enables a hardware agnostic approach which combined with a cache timing based techniques, yields a portable, memory safe system running wherever Rust is supported. The tiling strategy is applied to a subset of the polyhedral model, where loop nestings are tiled based on both the derived memory hierarchy and the observed data footprint per iteration.

\end{abstract}
\vspace{1em}
]

\section{Introduction}
\label{sec:introduction}
Many programs when observed in terms of their control flow behavior, are effectively expressible as straight line programs. This distinction is possible when the program execution pattern is a constant and independent of the user input passed through the given function's arguments. These straight line programs can be expressed as affine programs where an affine program is one where the control flow and memory access patterns can be statically predicted and analyzed, because they are determined entirely by linear arithmetic on loop indices and parameters known at compile time. This affine structure enables advanced mid-end compiler techniques to be performed in tractable time which can not be said for programs which have an unknown or possible infinite quantity of possible execution paths. The optimization which this paper will focus on is that of loop tiling. Tiling is the process in which additional loops are added to further nest a loop or a nesting of loops. A loop nesting is where a loop is placed within another loop which can continue and be arbitrarily deep. 

Tiling serves to improve data locality and enable better use of the memory hierarchy by partitioning the iteration space into smaller blocks that fit more effectively into cache. It acts as a method of reordering the execution of nested loop program sections to maximize spatial and temporal locality with respect to data traversal. Ideally, three levels of tiling are used, with tile sizes chosen to target the capacities of L1, L2, and L3 caches respectively. This hierarchical approach to locality is crucial because the cost of data movement is significantly higher than that of computation. Fetching data from main memory is orders of magnitude slower than accessing it from cache, making data movement the dominant factor in performance for many programs. By minimizing these costly memory transfers, tiling reduces the effective data movement complexity of a program, allowing compute resources to be better utilized with greater compute bandwidth.

The programs that benefit most from tiling are those with regular, predictable and affine programs. These include a broad class of applications in scientific computing, such as stencil computations, matrix multiplication, and others where array accesses and loop bounds are defined by linear organizations of loop indices. Similarly, many machine learning and AI workloads particularly those involving dense linear algebra, tensor operations, and convolutional neural networks exhibit affine access patterns and deeply nested loop structures making them ideal candidates for tiling. In these domains, large volumes of data must be processed repeatedly in structured ways, and performance is often bottlenecked by memory bandwidth and latency rather than raw computation. Tiling enables these programs to better exploit the memory hierarchy, significantly improving throughput by reducing cache misses and enhancing data reuse.

To optimize performance on modern architectures, many compilers optimizations on affine programs rely on auto tuning. Auto tuning is the process of automatically searching over a set of possible program configurations which includes tile sizes, unrolling factors, vectorization strategies, and more to identify the combination that yields the best performance for a specific hardware target. This search is typically done by running each version of the program configuration and recording the measured performance. While auto-tuning can produce highly optimized code tailored to a given machine, it is computationally expensive and time consuming, sometimes requiring hours or even days of profiling to converge on an optimal configuration. This makes it impractical in contexts where rapid compilation and deployment are required, or in cases where the cost of the overhead is not amortized.

Compared to the previous work of auto-tuning, latency based tiling performs tiling in a new way. The main contributions from this work are as follows:

\begin{itemize}
    \item \textbf{A novel compile time tiling strategy}: Latency Based Tiling, a method that derives tile sizes based on system level memory latency measurements instead of empirical auto-tuning. The latency based method constructs then interprets a miss ratio scale as a representation of a systems memory hierarchy which can be used portably for any compilation of a affine programs.
    
    \item \textbf{Triangular loop based latency profiling}: We use triangular loop structures to empirically model miss ratio scaling, effectively identifying the boundaries of L1, L2, and L3 cache levels with minimal interference from hardware prefetching. This prevention of prefetching makes the limitations of the hardware move obvious which is desired when caching particularly in a defensive manner. The latency profiling still incurs variance inherent to non-deterministic hardware and kernel constructs. The profiling also captures the defensive nature of running the measurements directly on a multi process environment which helps the accuracy of finding real world memory hierarchy rather than the idealized cache configuration which we would derive from looking up the machine specifications.
    
    \item \textbf{Hardware agnostic performance modeling}: Our method provides a one time profiling mechanism that generalizes across architectures, requiring no prior knowledge of cache sizes or vendor specific tuning. This means that the process can be ran on virtually any machine and is not tied down to vendor provided specifications.
    
    \item \textbf{Fast compile times with competitive performance}: Unlike auto-tuning, which requires extensive runtime benchmarking, our approach introduces negligible compile time overhead while achieving near optimal locality. This distinction provides a one time static overhead for a given machine rather than running for each compilation. This means that compilation with a latency based tiling model should incur an imperceptible increase in compile time whereas with auto-tuning compile time goes from a few seconds to a few hours and in some cases longer.
    
    \item \textbf{Safe and portable implementation in Rust}: We implement our system using safe Rust, which offers a rare combination of performance, memory safety, and portability. Rust achieves C like speed through zero cost abstractions and fine grained control over data layout and computation, making it ideal for performance critical tasks like loop tiling. Its ownership and borrowing model ensures memory safety at compile time without garbage collection, eliminating bugs such as use after free or data races while maintaining predictable performance. Additionally, Rust is highly portable across architectures from desktops to embedded systems thanks to robust cross compilation tools and platform agnostic design. This makes it a powerful and reliable choice for building efficient, safe, and widely deployable system level optimization tools.
    
    \item \textbf{Applicability to affine programs in key domains}: We demonstrate that Latency Based Tiling is well suited to affine programs, which are prevalent in domains such as scientific computing and AI workloads. These programs typically feature regular loop nests and predictable memory access patterns, such as those found in matrix operations, stencil computations, and tensor contractions. Because their control flow and data access are statically analyzable, affine programs enable effective compile time transformations like tiling. In both scientific and AI applications, performance is often limited not by compute throughput but by memory bandwidth and latency. By aligning tile sizes with the underlying memory hierarchy, our approach significantly improves cache utilization and data reuse, making these programs more efficient and scalable across architectures which include a convention L1, L2, and L3 memory hierarchy.
\end{itemize}

The proposed approach does however comes with several limitations. This in turn makes it the case that replacing auto tuning with latency based tiling in all cases would not be possible. Latency based tiling does not suit all use cases making it possible for there to be instances where auto tuning is the better route to be taken. While the latency based tiling approach demonstrates significant advancements in both compiler and program runtime, these gains face certain limitations associated with our approach:
\begin{itemize}
\item \textbf{Lack of low level code generation features}: The current implementation does not include loop unrolling or hardware specific vectorization. While these optimizations could be added as separate compiler passes, they have not yet been integrated or evaluated in this work. These optimizations are also not directly relevant to improvements in data movement complexity. Loop unrolling reduces the number of instructions executed by decreasing the number of control flow operations and better enabling out-of-order execution of a larger number of in flight instructions. The downside of loop unrolling is that it increases the size of the instruction stream, which can negatively impact instruction cache locality. Hardware specific vectorization particularly through short vector intrinsics or SVI enables massive data parallelism by allowing a single instruction to operate on multiple data elements simultaneously. However, vector based computations have the downside of requiring data to be aligned and structured in a way that supports vectorization, which may increase the complexity of data layout transformations. Additionally, their effectiveness is highly hardware dependent, meaning performance gains may not generalize well across different target architectures.

\item \textbf{Tradeoff between compile time and peak performance}: Latency based tiling emphasizes fast compilation and hardware agnostic portability, making it ideal for rapid iteration and deployment across diverse platforms. However, in performance critical environments where runtime efficiency is most important and compilation overhead is heavily amortized, such as HPC workloads (e.g., climate modeling or fluid dynamics) or large scale machine learning inference systems, the modest performance trade offs of latency based tiling may be unacceptable. In these settings, auto tuning despite its significant compile time cost can deliver highly optimized code by exhaustively exploring low level configuration spaces, leading to crucial performance gains that scale across massive compute workloads or reduce operational costs at hyperscale. The current solution also is not done with an automatic compilation system but this is something in the future I hope to include which should not be too demanding of a mid-end compilation pass to develop for affine programs with conventional loop nesting structures.

\item \textbf{Non-determinism in latency measurement}: Hardware and software systems are inherently non-deterministic due to a variety of uncontrollable and unpredictable factors. On modern machines, background processes, OS scheduling decisions, specialized and hidden replacement policies, interrupt handling, and prefetching behavior can all influence execution time in subtle ways. Additionally, cache contention from other processes or system level tasks can cause variance in latency, and out-of-order execution and speculative execution can obscure precise timing behavior \citet{kandemir1998}. These sources of variability make single run measurements unreliable for detecting consistent performance trends such as cache boundary transitions. To mitigate this, measurements are typically run multiple times and averaged to reduce noise and expose stable patterns. However, on machines subject to thermal throttling where the CPU reduces frequency to stay within safe temperature limits later iterations may run more slowly due to accumulated heat, thereby skewing averages and reducing measurement fidelity. In such environments, averaging may not fully compensate for changes in performance, making it difficult to obtain clean, repeatable latency curves.

\item \textbf{Prefetching interference}: Modern CPUs use hardware prefetchers that try to predict what data a program will need next and load it into the cache before it’s actually requested. This is done by detecting patterns in memory access such as cyclic or sawtooth preloading future addresses. Prefetching is generally beneficial for reducing perceived memory latency and improving throughput, especially in memory bound programs, because it hides the delay of fetching data from slower levels of the memory hierarchy. However, for our purposes where I aim to precisely measure latency and identify cache boundaries prefetching can be problematic. It artificially smooths over memory access times, making it harder to detect where true cache capacity limits are. For example, even when data no longer fits in cache, the prefetcher may still deliver it by preemptively evicting other data, skewing the latency measurements and distorting the miss ratio curve. In such cases, disabling or minimizing prefetching can yield more accurate insights into memory hierarchy behavior and data movement complexity. While disabling prefetching is not used in this work latency based tiling does minimize the prefetching to some extent.

\item \textbf{Assumption of triangular loop pattern}: The latency model assumes that triangular loops accurately reflect the typical access patterns of target programs. However, these synthetic access patterns may not match the complexity of real world loops, potentially reducing the accuracy of the derived tiling sizes. There may be cases where the structure of a loop nesting is affine but is not correctly modeled by the approximation of triangular loop. The subset of polybench used in the evaluation of latency based tiling avoids this issue.

\item \textbf{Limited support for dynamic or irregular programs}: The method assumes affine access patterns and predictable memory behavior. Programs with input dependent control flow, dynamic data structures, or indirect indexing may not benefit from this tiling strategy and could require more adaptive or hybrid approaches. This limitation is in no way trivial and is also faced by all polyhedral based systems.
\end{itemize}

\section{Background}
\label{sec:background}

\subsection{Affine Programs and Loop Nests}
Affine programs are a class of computations where loop bounds and memory access patterns can be expressed as linear or affine functions of loop indices and compile time constants. These programs are characterized by regular, statically analyzable control flow and data access patterns, making them amenable to powerful compile time optimizations in a provably tractable amount of time. Common examples include dense linear algebra namely matrix multiplication, stencil computations, and tensor operations prevalent in scientific computing and machine learning. The polyhedral model is a mathematical framework for analyzing and transforming loop nests which is particularly effective for optimizing such programs, as it enables precise dependency analysis and loop transformations like tiling, fusion, and more.

\subsection{Memory Hierarchy and Data Locality}

Modern architectures employ a hierarchical memory system comprising registers, L1/L2/L3 caches, and main memory to mitigate the growing performance gap between processor speed and memory latency. The effectiveness of this hierarchy depends on data locality which is the reuse of cached data to minimize expensive accesses to slower memory levels.

\textit{Temporal locality} occurs when the same data is accessed repeatedly within a short time window, while \textit{spatial locality} arises when nearby data (e.g., array elements in contiguous memory) are accessed in sequence. Loop tiling improves both forms of locality by partitioning iteration spaces into smaller blocks or tiles that fit into cache, reducing capacity misses and bandwidth pressure.

Each level of the memory hierarchy comes with increasing capacity but also increasing latency. While something like addition takes 1 cycle and multiplication may take 3 to 5 cycles typically but a typical cache access may take significantly longer and only grow with every increasing level in the memory hierarchy. Accessing main memory can take a thousand cycles, while access to disk or virtual memory in the case of a page fault can cost millions of cycles. These dramatic differences in latency underscore the importance of keeping working sets within the faster levels of cache.

In multi core, multi process systems, caches are often shared among cores, especially the L3 cache, which is typically shared across all cores on a chip. This shared nature introduces interference: when multiple cores or processes compete for the same cache resources, the effective cache size available to each thread is reduced, leading to increased cache misses and degraded performance. This variability makes it difficult to reason about memory behavior using static models or vendor reported cache sizes, motivating the need for empirical profiling techniques like the one described in this work.

\subsection{Traditional Tiling Strategies}

Loop tiling improves cache efficiency by dividing loop iteration spaces into smaller, cache friendly memory segments better conforming to the ideal data movement ordering. This is typically implemented by restructuring loop nests to place the nesting within outer tile or control loops which iterate over tiles where the bounds of the loops which were there before stay the same but now the bounds are defined by the new outer nesting which also presents itself in the actual data accesses as single value indexes become multiple value indexes taking both the existing loops and newly introduced tile loop dependencies into account. The effectiveness of tiling heavily depends on choosing an appropriate tile size:
\begin{itemize}
    \item If the tile is too small, overhead from additional loop control and reduced amortization of loop body work can outweigh cache benefits. A simple tiling solution which aims to make conservative optimizations would likely make this mistake.
    \item If the tile is too large, it may exceed the available cache capacity, leading to frequent cache evictions and performance degradation due to capacity and conflict misses. This would remove all desired locality with respect to a given level of cache and looses any of the value of performing the tiling. This can often be the case when taking the cache sizes directly without accounting for the needed awareness of a defensive strategies.
\end{itemize}

Two conventional strategies for selecting tile sizes are commonly used:

\begin{itemize}
    \item \textbf{Static analytical models}: Static analytical models rely on compile time knowledge of hardware parameters such as cache size, line size, associativity, and memory latency to compute a theoretically optimal tile size. These models often analyze data reuse patterns, loop bounds, and access strides to estimate the working set size and predict cache behavior \citet{falcon2024}. While analytically they are clean and seemingly reflective, static models struggle to capture dynamic hardware behaviors like speculative execution, prefetching, cache sharing in multi-core systems, and memory level parallelism. As a result, they may yield suboptimal tile sizes in practice, particularly on modern, deep memory hierarchies. These static analyses can often produce tiles which are too large which is usually much worse that a tile which is slightly to small. Latency based tiling serves in a sense as a method to dynamically under approximate caches for a better real world representation of caches and system behavior.
    
    \item \textbf{Auto-tuning}: This approach treats tile size selection as a state space search problem over a set of possible configurations. A set of candidate tile sizes is generated, and performance is measured for each variant by compiling and executing them on the target hardware. The best performing configuration is then chosen. Auto-tuning is extendable to a wide variety of hardware as it does not require information about the target system and informs itself through the latency information. Through an expansive search space it finds the highest performing variant yielding near optimal results. However, it incurs significant cost in both time and compute resources, particularly when the tuning space is large or when rapid deployment is required. This issue of extremely long run times are alleviated by the latency based tiling solution but the auto-tuning still provides the best results to the point where auto-tuning can in a sense be considered the oracle of optimal tiling derivation.
\end{itemize}

\subsection{Challenges in Memory Latency Measurement}
Accurately profiling memory hierarchy behavior is complicated by:
\begin{itemize}
    \item \textbf{Hardware prefetching}: Modern CPUs predict and preload data, obscuring true cache miss penalties.
    \item \textbf{Non-determinism}: Background processes, interrupts, and OS scheduling introduce noise in latency measurements.
    \item \textbf{Shared resources}: In multi-core systems, caches are shared across processes, causing unpredictable contention.
\end{itemize}
Latency Based Tiling addresses these challenges by using \textbf{triangular loops} which serves a synthetic workload designed to minimize prefetcher interference in order to empirically measure miss ratios and infer cache boundaries without relying on vendor specifications. Latency Based Tiling running passes multiple times and takes the average to also help prevent non-deterministic spikes to completely reflect a given point in the MRS. 

\subsection{Rust for Performance-Critical Systems}
Rust's combination of zero cost abstractions, fine grained memory control, and compile time safety makes it uniquely suited for implementing system level optimizations. Its ownership model statically enforces aliasing and lifetime rules, eliminating entire classes of memory errors such as use after free, double free, and dangling pointers without requiring garbage collection. Unlike C or C++, memory safety in Rust is guaranteed and enforced by the compiler through its strongly typed type system. The portability of Rust binaries across platforms (e.g., x86, ARM) supports the hardware agnostic aspirations of Latency Based Tiling. Rust's growing ecosystem and compiler support for cross compilation enable developers to target a wide range of devices without rewriting system level code.

\subsection{Defensive Tiling in Shared Environments}
In multi-process systems, cache contention from unrelated workloads can degrade performance. Defensive tiling which is a strategy that intentionally under-approximates cache sizes to reserve bandwidth for CPU sharing over multiple processes is critical for robust performance. Latency Based Tiling inherently adopts this approach by deriving cache sizes from observed latency transitions, which naturally reflect shared resource overheads.

\section{Deriving Latency Based Tiling For Realistic Performance}
\label{sec:implementation}

The system derives cache hierarchy parameters through empirical latency measurement using synthetic triangular access patterns. These patterns create non sequential memory access streams that defeat hardware prefetching while maintaining predictable stride patterns that expose true cache capacity limits.

\subsection{Core Measurement Strategy}
The implementation focuses on two key memory access patterns:

\begin{itemize}
\item \textbf{Cyclic (Forward-Forward) A B C A B C}

\item \textbf{Sawtooth (Forward-Backward) A B C C B A} 
\end{itemize}

\begin{figure}[t]
\begin{minipage}{\linewidth}
\begin{verbatim}
// Triangular access pattern generator
fn triangular_access(size: usize) -> f64 {
    let mut buffer = vec![0u64; size];
    let mut p = 0;
    let mut stride = 0;
    
    // Initialize pattern
    while p < size {
        let idx = (stride + p) % size;
        buffer[idx] = idx as u64;
        stride += 8; // Non power of two stride
        p += 8;
    }

    // Time pattern execution
    let start = Instant::now();
    let mut access: usize = 0;
    for _ in 0..size {
        access = buffer[access] as usize;
    }
    start.elapsed().as_secs_f64()
}
\end{verbatim}
\end{minipage}
\caption{Key measurement pattern implementation in Rust}
\label{fig:code}
\end{figure}

\subsection{Implementation Details}
The measurement process (Figure~\ref{fig:code}) operates through:

\begin{enumerate}
\item \textbf{Pattern Generation}: Lines 4-10 create triangular access patterns using non power of two strides to defeat prefetchers

\item \textbf{Timed Execution}: Lines 13-17 measure end-to-end latency for progressively larger buffers

\item \textbf{Statistical Analysis}: Identifies cache boundaries where latency increases exceed thresholds
\end{enumerate}

Key advantages of our Rust implementation:
\begin{itemize}
\item Memory safety through compile time checks
\item Portable timing via \texttt{std::time::Instant}
\item Deterministic patterns through controlled strides
\end{itemize}

\section{Latency Analysis and Cache Boundary Detection}
\label{sec:latency-analysis}

The system employs a robust statistical approach to identify cache hierarchy boundaries from raw latency measurements. This section details the data representation and analysis pipeline that transforms timing data into actionable cache size estimates.

\subsection{Data Representation}
The measurement pipeline uses Rust's strong type system to maintain data integrity throughout analysis:

\begin{figure}[h]
\begin{minipage}{\linewidth}
\begin{verbatim}
/// Represents a single latency measurement point
#[derive(Serialize, Deserialize, Clone)]
pub struct LatencyData {
    pub size: usize,  // Tested working set size (bytes)
    pub avg_time: f64,// Normalized access latency (ns)
}

/// Aggregated results for a memory access pattern
#[derive(Serialize, Deserialize)]
pub struct PatternResults {
    pub peaks: (usize, usize, usize),// All peaks
    pub data: Vec<LatencyData>,      // All measurements
}
\end{verbatim}
\end{minipage}
\caption{Core data structures for latency analysis}
\label{fig:data-struct}
\end{figure}

The design (Figure~\ref{fig:data-struct}) provides three key advantages:

\begin{itemize}
    \item \textbf{Immutability by Default}: The \texttt{Clone} derivation enables safe data sharing while preventing accidental mutation
    \vspace{-2.5em}
    \item \textbf{Memory Efficiency}: The \texttt{Vec} storage provides cache friendly contiguous allocation
    \vspace{-2.5em}
    \item \textbf{Self Contained Results}: Tuple stored peaks with confidence scores enable standalone analysis
\end{itemize}

\subsection{Boundary Detection Algorithm}
The core algorithm identifies cache boundaries through statistical analysis of latency discontinuities:

\begin{algorithm}[H]
\caption{Cache Boundary Detection}
\begin{algorithmic}[1]
\Procedure{FindBoundaries}{$data$}
    \State $diffs \gets \text{empty vector}$
    \For{$window \in data.\text{windows}(2)$}
        \State $\Delta size \gets window[1].size - window[0].size$
        \State $\Delta time \gets window[1].avg\_time - window[0].avg\_time$
        \State $deriv \gets \Delta time / \Delta size$
        \State $diffs.\text{push}((window[1].size, deriv))$
    \EndFor
    
    \State \text{Sort $diffs$ by $deriv$ descending}
    \State $peaks \gets \text{Top 3 } diffs$
    \State \text{Sort $peaks$ by $size$ ascending}
    
    \State $confidence \gets [\text{normalize}(peaks[0..2].deriv)]$
    \State \Return $(peaks, confidence)$
\EndProcedure
\end{algorithmic}
\end{algorithm}

\subsubsection{Differential Analysis}
The discrete derivative of the latency curve is computed by:
\begin{equation}
\frac{\Delta \text{time}}{\Delta \text{size}} = \frac{t_{i+1} - t_i}{s_{i+1} - s_i}
\end{equation}

This transformation converts absolute latency measurements into a sensitivity function where sharp increases indicate cache capacity boundaries.

\subsubsection{Statistical Ranking}
The derivative values are sorted by magnitude to identify the most significant jumps. Rust's safety is guaranteed through:

\begin{itemize}
    \item Bounded float comparisons via \texttt{partial\_cmp}
    \item Stable sorting to preserve measurement ordering
    \item Guardrails against NaN values
\end{itemize}

\subsubsection{Confidence Scoring}
Each detected boundary receives a confidence score:
\begin{equation}
c_i = \frac{d_i}{\sum_{j=0}^2 d_j}
\end{equation}

where $d_i$ represents the normalized derivative magnitude at boundary $i$.

\section{Experimental Evaluation}
\label{sec:evaluation}

The effectiveness of Latency Based Tiling is evaluated across eleven benchmarks from the PolyBench suite. All experiments were conducted on an Apple M2 MacBook Air with 8 CPU cores and 16GB of unified memory, running macOS. Benchmarks were compiled using the Rust compiler with \texttt{opt-level = 0} to disable compiler optimizations. This ensured that the executed code closely matched the written source, avoiding unintended performance improvements that could obscure the impact of tiling transformations.

\begin{table}[ht!]
\centering
\small
\begin{tabular}{lrrr}
\toprule
\textbf{Benchmark} & \textbf{Tiled (s)} & \textbf{Non Tiled (s)} & \textbf{Speedup} \\
\midrule
3mm         & 3.755 & 4.561 & 1.21x \\
2mm         & 2.696 & 2.901 & 1.08x \\
gemm        & 1.304 & 1.498 & 1.15x \\
syrk        & 0.870 & 0.909 & 1.04x \\
covariance  & 0.605 & 0.591 & 0.98x \\
doitgen     & 0.207 & 0.217 & 1.05x \\
seidel-2d   & 0.164 & 0.172 & 1.05x \\
bicg        & 0.059 & 0.059 & 1.00x \\
fdtd-2d     & 0.055 & 0.057 & 1.03x \\
atax        & 0.018 & 0.018 & 1.00x \\
jacobi-2d   & 0.014 & 0.014 & 0.97x \\
\bottomrule
\end{tabular}
\caption{Execution time comparison between tiled and non tiled implementations (lower is better).}
\label{tab:performance-results}
\end{table}

\begin{figure}[ht!]
    \centering
    \includegraphics[width=\linewidth]{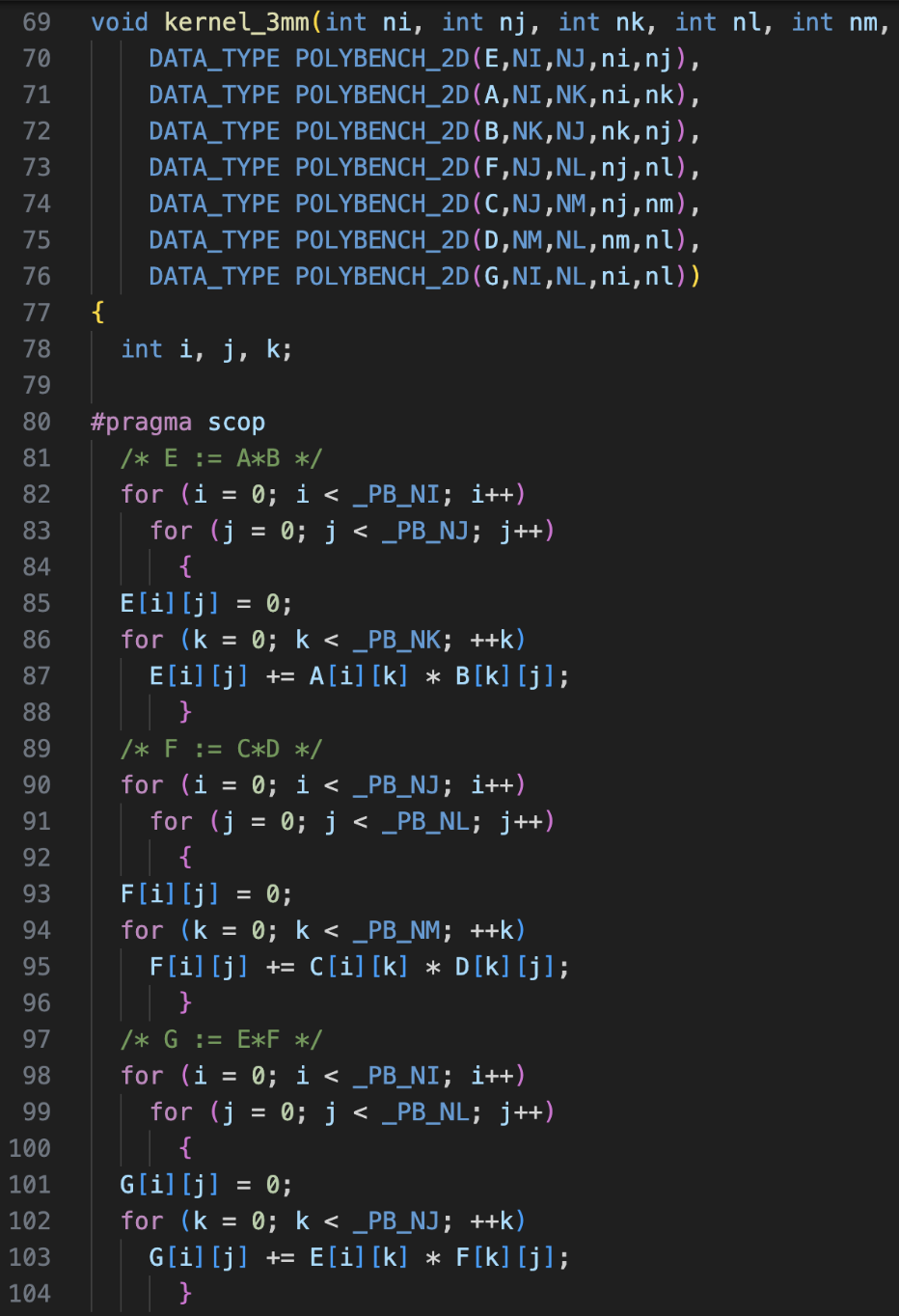}
    \caption{Section of 3mm without tiling insertion.}
    \label{fig:3mm-untiled}
\end{figure}

\begin{figure}[ht!]
    \centering
    \includegraphics[width=\linewidth]{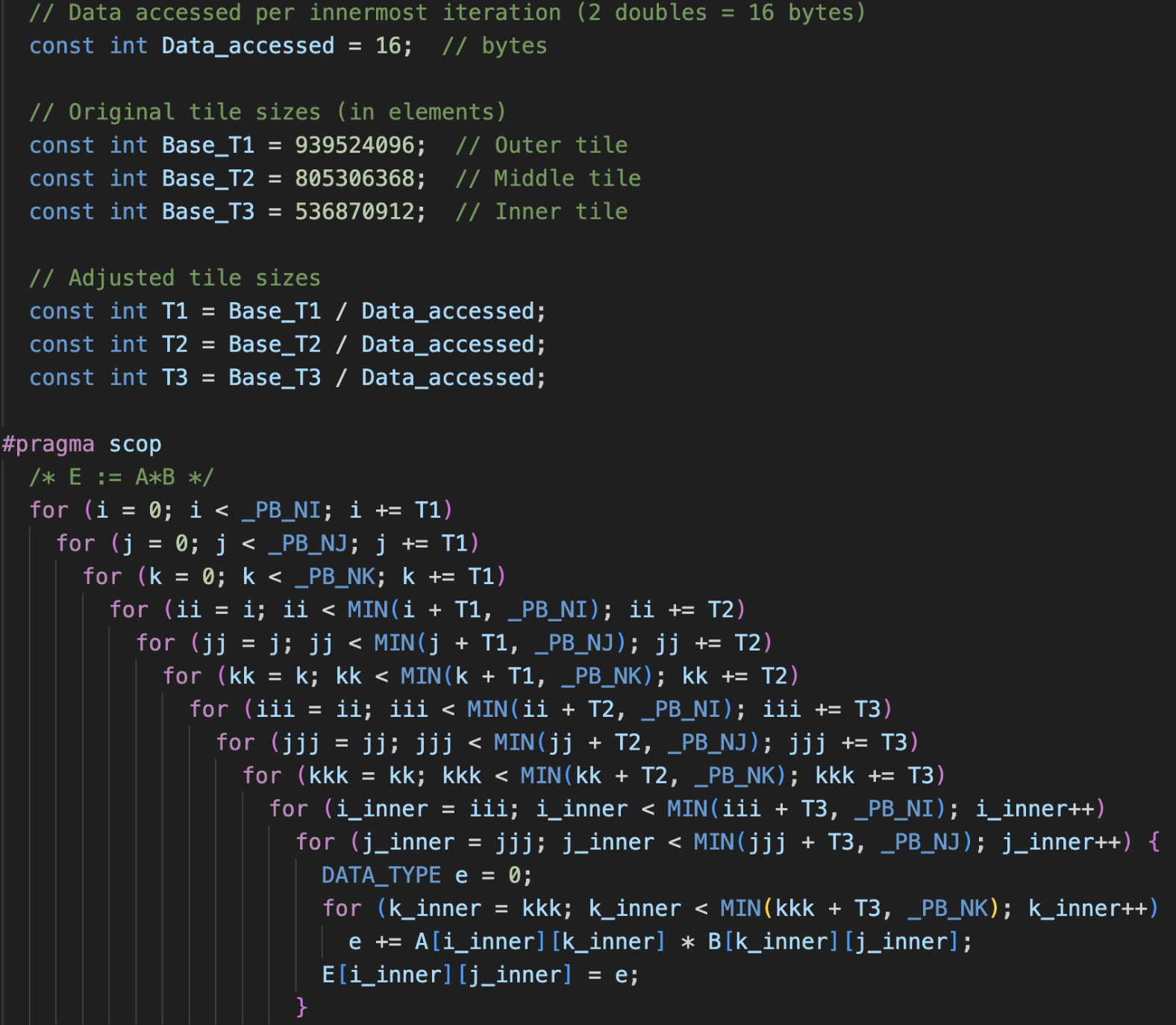}
    \caption{Section of 3mm with tiling insertion.}
    \label{fig:3mm-tiled}
\end{figure}

\subsection{Performance Analysis}
The results demonstrate three distinct performance patterns:

\begin{itemize}
\item \textbf{Matrix Operations (3mm, 2mm, gemm)}: Achieved consistent speedups (1.08-1.21x) due to improved cache utilization in these compute bound kernels. The 3mm benchmark shows the greatest benefit (21\% faster) as it involves chained matrix multiplications that particularly benefit from tiled memory access patterns.

\item \textbf{Moderate Improvements (syrk, doitgen, seidel-2d)}: Showed modest 5\% speedups from tiling. These benchmarks have more irregular access patterns that partially benefit from improved locality but are less computationally intensive.

\item \textbf{Neutral/Negative Impact (covariance, atax, jacobi-2d, bigcg)}: Demonstrated negligible differences or slight regressions. These memory bound benchmarks have working sets that either fit entirely in cache or exhibit strided accesses that don't benefit from tiling. The 0-3\% overhead comes from additional loop control instructions.
\end{itemize}

\begin{figure}[ht!]
    \centering
    \includegraphics[width=\linewidth]{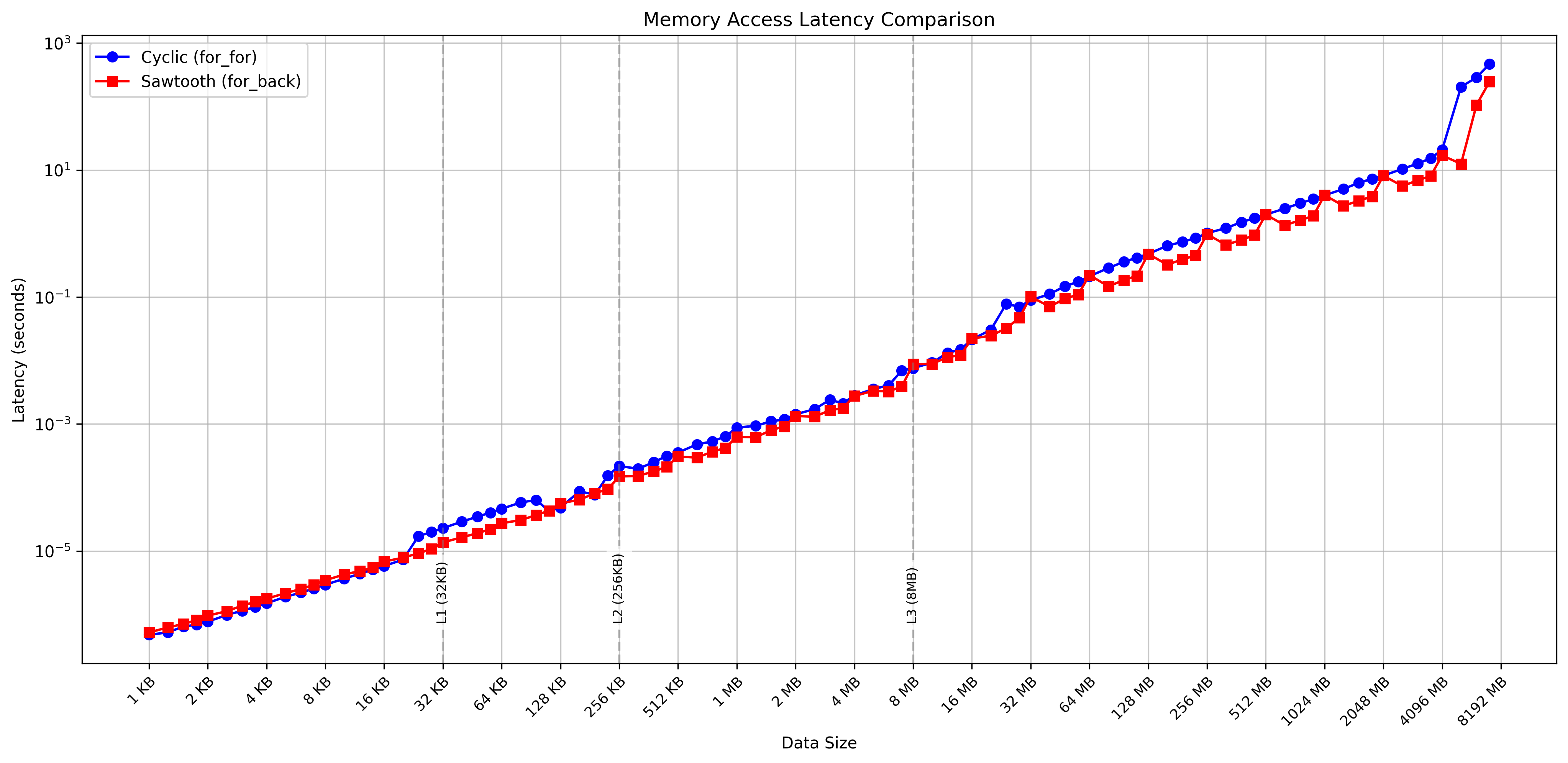}
    \caption{Data Traversal Latency with Approximate Cache Levels}
    \label{fig:latency-comparison}
\end{figure}

\begin{figure}[ht!]
    \centering
    \includegraphics[width=\linewidth]{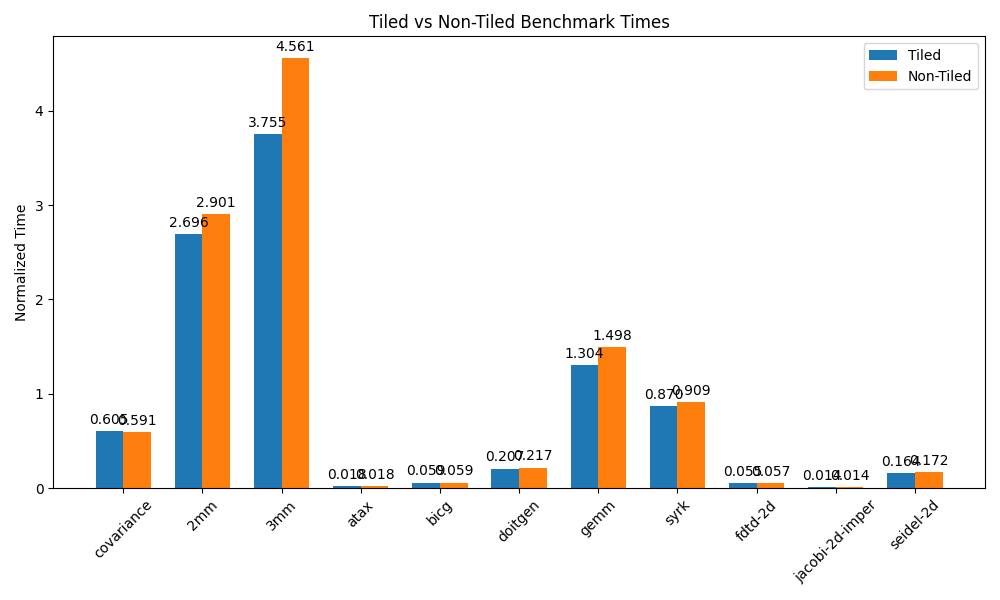}
    \caption{Relative Benchmark times (Lower is better)}
    \label{fig:speedup-duplicate}
\end{figure}

\begin{figure}[ht!]
    \centering
    \includegraphics[width=\linewidth]{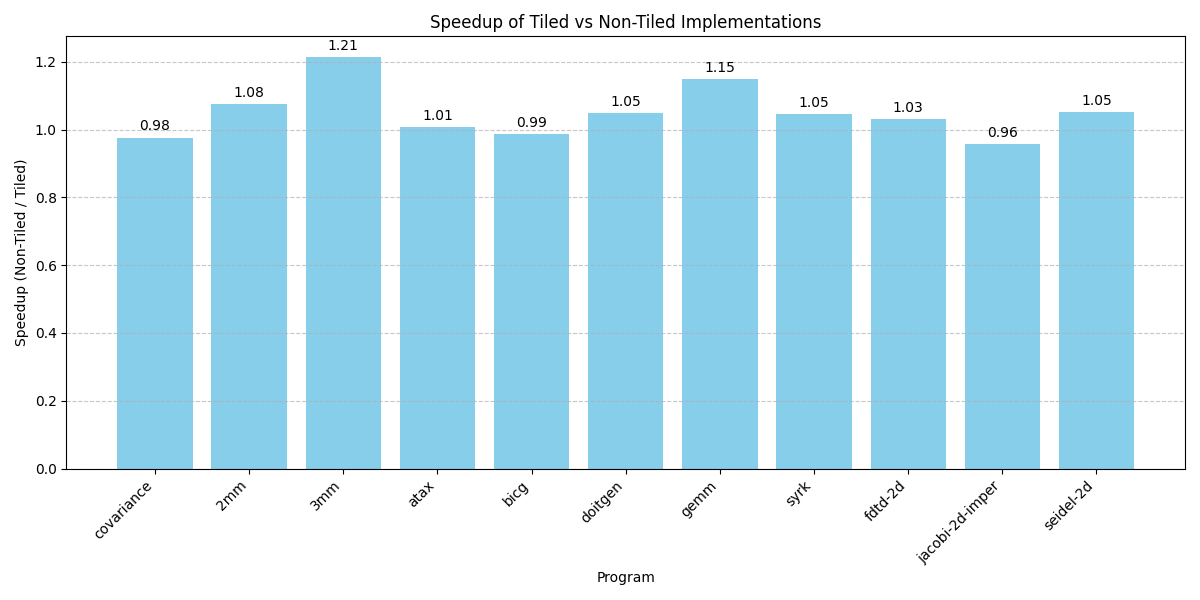}
    \caption{Ratio between non tiled/tiled run time (Higher is better)}
    \label{fig:cache-miss-bar}
\end{figure}

\subsection{Compile Time Overhead}
The latency based tiling approach added negligible compilation time and is not even worth measuring at this point in the development of latency based tiling. The tiling of the selected programs has been performed by hand and manually insertion of the tile values was done. This makes compiler measurements not worthwhile but it can still be clearly asserted that this optimization technique would by far in a way be the closest in compile time to compilation with a expected but minimal degree of optimization.

\section{Conclusion}
\label{sec:conclusion}
The work I have presented is Latency Based Tiling, a novel way of performing tiling which captures the memory hierarchy of a system and in turn enables tiling in a way which is significantly faster than auto-tuning in terms of compilation and ideally marginally slower. We show that when this system is used there is upwards of a 21 percent performance increase which given further improvements I expect to improve. It is also only results captured on a personal device but more advance and common environments for affine program execution would likely see a greater performance uplift.

I strongly believe that this work can be furthered. Firstly the latency measurement process can better explore the search space for which the latency increases are found, and can also apply artificial bounds on how large or small the given cache levels can be and have individual search within those spaces to get more effective results and at a faster run time. I also need to develop the automatic mid-end compilation pass rather than performing hand written tiling on the programs makes the approach more scalable and effective as a real world tool. I would also like to see about adding unrolling and vectorization as an additional compilation pass which can be called directly as an argument. Additionally I plan on bringing the my work of LALA or Loop Asymptotic Locality Analysis to the compilation process enabling the fastest and most performant loop interchange possible. All of these improvements and more can be added and will further demonstrate the novelty and benifits of Latency Based Tiling.

\bibliographystyle{plainnat}

\end{document}